# Sensing and Link Model for Wireless Sensor Network: Coverage and Connectivity Analysis

Ashraf Hossain
Department of Electronics & Communication Engineering
NIT Silchar, Silchar, Assam –788010, India
E-mail: hossain_ashraf@rediffmail.com

Rashmita Mishra
Department of Electronics & Communication Engineering
BPPIMT, Kolkata, India
E-mail: rashmi_misha@yahoo.co.in

*Abstract*—**Coverage and connectivity both are important in wireless sensor network (WSN). Coverage means how well an area of interest is being monitored by the deployed network. It depends on sensing model that has been used to design the network model. Connectivity ensures the establishment of a wireless link between two nodes. A link model studies the connectivity between two nodes. The probability of establishing a wireless link between two nodes is a probabilistic phenomenon. The connectivity between two nodes plays an important role in the determination of network connectivity. In this paper, we investigate the impact of sensing model of nodes on the network coverage. Also, we investigate the dependency of the connectivity and coverage on the shadow fading parameters. It has been observed that shadowing effect reduces the network coverage while it enhances connectivity in a multi-hop wireless network.**

## I. INTRODUCTION

Wireless sensor network consists of a large number of energy-constrained nodes that are deployed for monitoring multiple phenomena of interest [1]. Nodes are randomly placed one by one. Any event is said to be detectable if at least one node lies within its observable range. Coverage has been studied for WSN by several authors [2–9]. All the previous works consider only the Boolean sensing model. Tsai [5] has studied sensing coverage for randomly deployed wireless sensor network in shadow fading environment. In [5] the Boolean sensing model and shadow fading sensing model have been used for analyzing network coverage. Besides these there is another reported sensing model in the literature known as Elfes sensing model [6].

Connectivity for multi-hop network has been studied in [10] for shadow-fading environment. In this paper we have again presented connectivity analysis to get insight how fading parameters affect the network coverage and connectivity of the network.

The rest of the paper is organized as follows. Section II presents the system model and sensing model. Section III provides network coverage. Section IV provides link model and connectivity analysis. Section V presents numerical results for performance studies of WSN. Finally, section VI concludes the paper.

## II. SYSTEM MODEL AND SENSING MODEL

We consider an area of interest $A$ where $N$ nodes are randomly distributed. Let us assume that the nodes are uniformly deployed with homogeneous node density $\rho = N/A$. All nodes are capable of performing sensing, receiving and transmitting.

Two detection models are reported: individual detection model and cooperative detection model [3]. For simplicity we assume the first one where each sensor independently detects an event.

In the individual detection model, a node detects an event if the received signal strength is greater than the threshold value of detection, known as the sensing sensitivity. The detection process depends on the strength of the emitted signal, behavior of the environment and the hardware of the node. In the literature the most cited sensing model is the Boolean sensing model [2, 3].

### A. Boolean sensing model

It is very simple one. If the occurrence of the event is within the sensing range of a node then the event will be assumed to be detected, otherwise not (Fig. 1(a)). This model ignores the dependency of the condition of the environment and the strength of the emitted signal on the task of sensing.

### B. Shadow fading sensing model

The dependency of all the factors has been taken into account in the shadow fading sensing model. Here, the sensing ability of a node is not uniform in all the directions (Fig. 1(b)). This is similar to shadowing in radio wave propagation. A detailed analysis is given in [5].

Let $x$ be the distance between an arbitrary sensor node and an event of interest. We assume that the power emitted by the event is equal to $P_s$.

Assuming log-normal shadowing path loss model, the received power at the node due to the event is

$$P(x)[dBm] = P_s[dBm] - PL(x)[dB] \qquad (1)$$

where path loss is

$$PL(x) = \overline{PL}(x_0) + 10n\log_{10}(x/x_0) + X_\sigma \qquad (2)$$

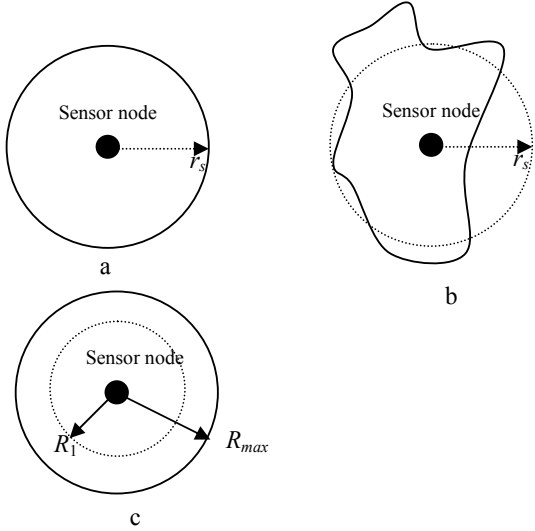

Fig. 1. Different sensing models: (a) Boolean sensing model, (b) Shadow fading sensing model, (c) Elfes sensing model.

The parameter $n$ is known as path loss exponent and $X_\sigma$ is a zero-mean Gaussian distributed random variable (in dB) with standard deviation $\sigma$ (in dB) [11].

Let $P_{s,th}$ be the sensitivity of the sensor node, then the transmission will be successful if $P(x) \geq P_{s,th}$. Thus the probability of detecting the event under shadow fading environment is

$$P_{det}(x) = \text{Prob}\{P(x) \geq P_{s,th}\} = Q\left(\frac{P_{s,th} - (P_s - \overline{PL}(x))}{\sigma}\right) \qquad (3)$$

where $\qquad \overline{PL}(x) = \overline{PL}(x_0) + 10n\log_{10}(x/x_0) \qquad (4)$

Let $r_s$ be the sensing range in non-shadowed environment for which (5) does hold well.

$$P(r_s) = P_{s,th} = P_s - \overline{PL}(x) - 10n\log_{10}(r_s/x) \qquad (5)$$

Plugging (5) into (3) yields

$$P_{det}(x) = Q\left(\frac{10n\log_{10}(x/r_s)}{\sigma}\right) \qquad (6)$$

where $\qquad Q(x) = \frac{1}{\sqrt{2\pi}} \int_x^\infty e^{-y^2/2} dy \qquad (7)$

*C. Elfes sensing model*

The Boolean sensing model is a deterministic model. The analysis for Shadowing sensing model in [5] reveals a probabilistic sensing model. Another probabilistic sensing model reported in the literature is Elfes sensing model.

According to Elfes sensing model [6], the probability that a sensor detects an event to a distance $x$ is

$$p(x) = \begin{cases} 1, & 0 \leq x \leq R_1 \\ e^{-\lambda(x-R_1)^\beta}, & R_1 < x < R_{max} \\ 0, & x \geq R_{max} \end{cases} \qquad (8)$$

where $R_1$ defines the starting of uncertainty in sensor detection and the parameters $\lambda$ and $\beta$ are adjusted according to the physical properties of the sensor. $R_{max}$ is the maximum sensing range of the node. This model is more general because it becomes Boolean sensing model when $R_1 = R_{max}$.

For $R_1 = 0, \beta = 1$, (8) becomes

$$\begin{aligned} p(x) &= e^{-\lambda x}, \ R_{max} > x > 0 \\ &= 0, \ x \geq R_{max} \end{aligned} \qquad (9)$$

Fig. 1(c) depicts two sensing zones in Elfes sensing model.

### III. NETWORK COVERAGE

*A. Network coverage for Boolean sensing model*

The sensing model has an impact on network coverage. Network coverage means how well an area is being monitored by a network. Network coverage depends on the sensing ability of a node, node number and the monitoring area. Network coverage or coverage fraction is defined as the ratio of the covered area to the total area of interest. A general calculation for coverage fraction is as follows.

Let $r_s$ be the sensing radius of the sensor. Any event in $A$ will be detected by any arbitrary sensor if it is within $r_s$ distance away from the event (Fig. 2). The probability that the event will be detected by the arbitrary sensor is

$$p = \frac{\pi r_s^2}{A} \qquad (10)$$

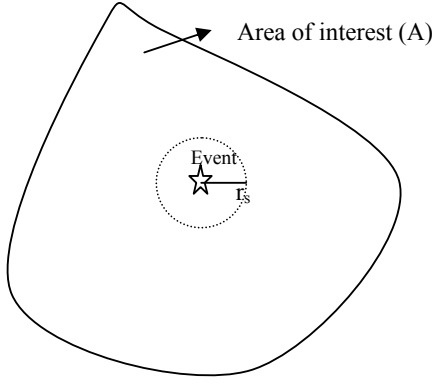

Fig. 2. An event occurs in an area of interest.

The probability that the event will not be detected by the arbitrary sensor is equal to $(1 - p)$. The number of nodes deployed randomly is $N$. Thus the probability that the event will not be detected by any one of the node is

$$P_{un\det} = (1-p)^N \quad (11)$$

The probability that the event will be detected by at least one node is known as the coverage fraction and is

$$f_a = 1 - P_{un\det} = 1 - (1-p)^N \quad (12)$$

Equation (12) can also be approximated as

$$f_a = 1 - e^{-Np} \quad (13)$$

Here, we have neglected the boundary effect. Also we have assumed Boolean sensing model in deriving the coverage fraction. The same for shadowing environment is given in [5]. Now, we present the network coverage in the light of Elfes sensing model.

### B. Network coverage for Elfes sensing model

We assume that nodes are randomly deployed over an area $A$ and one such sensor is at a distance $x$ from the event (Fig. 3). The probability that a specific node is deployed at a location with a distance $x$ to the event is $2\pi x dx/A$, where $dx$ is a small increment in distance $x$. The probability that the event is sensed by the sensor is

$$P_{\det} = \frac{1}{A} \int_{x=0}^{R_{\max}} p(x) 2\pi x dx$$
$$= \frac{2\pi}{A} \int_{x=0}^{R_{\max}} e^{-\lambda x} x dx$$
$$= \frac{2\pi}{A} \frac{1}{\lambda^2} \left[1 - e^{-\lambda R_{\max}}(1+\lambda R_{\max})\right] \quad (14)$$

where $A$ is the total area of interest.

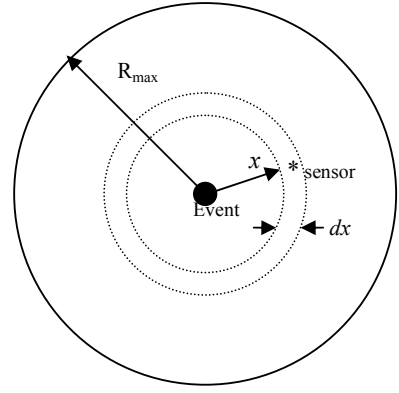

Fig. 3. The sensing for probabilistic model

The coverage fraction can be found from (12) or (13) where $p$ will be replaced by $P_{\det}$ and can be simplified as

$$f_a = 1 - \exp\left[-\frac{2\pi N}{A\lambda^2}\left\{1 - (\lambda R_{\max}+1)e^{-\lambda R_{\max}}\right\}\right] \quad (15)$$

Equation (15) provides the necessary formula for studying coverage based on probabilistic sensing model.

### C. Network coverage without approximation in Elfes sensing model

In the previous section, we have derived the coverage fraction for approximate Elfes sensing model. In this section we provide the coverage fraction for Elfes sensing model.

Using (8) we have the probability of detection an event by a node (Fig. 3)

$$P_{\det} = \frac{\pi R_1^2}{A} + \frac{1}{A} \int_{x=R_1}^{R_{\max}} p(x) 2\pi x dx \quad (16)$$

Substituting $\beta=1$ in (8), the simplified form of (16) becomes

$$P_{\det} = \frac{\pi R_1^2}{A} + \frac{2\pi}{A\lambda^2}\left[(1+\lambda R_1) - e^{-\lambda(R_{\max}-R_1)}(1+\lambda R_{\max})\right] \quad (17)$$

Similar to previous section the simplified expression for coverage fraction will be found using (17).

### D. Network coverage for shadow fading environment

The probability that an event will be detected is given by [5]

$$P_{\text{det}} = \frac{2\pi}{A}\int_0^\infty P_{\text{det}}(x)xdx$$
$$= \frac{2\pi}{A}\int_0^\infty Q\left(\frac{10n\log_{10}(x/r_s)}{\sigma}\right)xdx \quad (18)$$

The network coverage is dependent on node sensing model. The expression of it is very simple for Boolean sensing model. It is quite complex for probabilistic sensing model. It reveals from [5] that the network coverage for shadowing sensing model gets reduced as shadowing parameter increases.

## IV. LINK MODEL AND CONNECTIVITY ANALYSIS

Link model analysis is required for multi-hop wireless network such as wireless sensor network. We assume that a node has finite communication range, $R_0$. There are two link models: non-shadowed link model, and shadowed link model. In the first case, it is assumed that if two nodes lie within the communication range, then there is a link between them. This is also known as radio disk model. Fig. 4(a) depicts the non-shadowed model where the communication range is a deterministic parameter and independent of environments. However, it has been experienced in wireless communication that there is a certain probability of establishing a link between two nodes that are more than $R_0$ distance away from each other. Similarly, there is a probability of establishing no link between two nodes that are located within $R_0$ distance. Fig. 4(b) depicts the link model for shadowed environment.

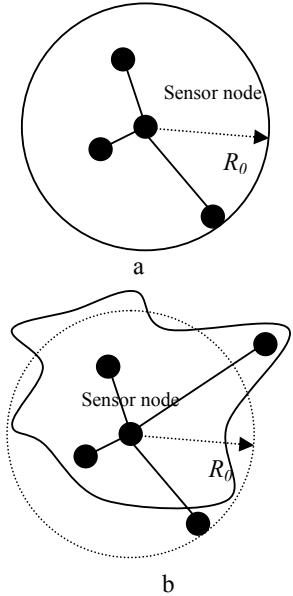

Fig. 4. Different link models: (a) Non-shadowed (radio disk) model, (b) Shadowed model.

*Probability of establishing a link between two nodes*

Let $d$ be the distance between two nodes $i$ and $j$. We assume that all the nodes are capable of transmitting equal power $P_t$. We would like to study the connectivity between the two nodes in a shadow fading environment for known distance of separation.

Assuming log-normal shadowing path loss model, the received power at node-$j$ is

$$P_j(d)[dBm] = P_t[dBm] - PL(d)[dB] \quad (19)$$

where path loss is

$$PL(d) = \overline{PL}(d_0) + 10n\log_{10}(d/d_0) + X_\sigma \quad (20)$$

The parameter $n$ is known as path loss exponent and $X_\sigma$ is a zero-mean Gaussian distributed random variable (in dB) with standard deviation $\sigma$ (in dB).

Let $P_{\text{rth}}$ be the sensitivity of the sensor node, then the transmission will be successful if $P_j(d) \geq P_{\text{rth}}$. Thus the probability of establishing a link is

$$\text{Prob}(\text{link is established}) = \text{Prob}\{P_j(d) \geq P_{\text{rth}}\}$$
$$= Q\left(\frac{P_{\text{rth}} - (P_t - \overline{PL}(d))}{\sigma}\right) \quad (21)$$

where $\overline{PL}(d) = \overline{PL}(d_0) + 10n\log_{10}(d/d_0) \quad (22)$

Let $R_0$ be the communication range in non-shadowed environment for which (23) does hold good.

$$P_j(R_0) = P_{\text{r}th} = P_t - \overline{PL}(d) - 10n\log_{10}(R_0/d) \quad (23)$$

Plugging (23) into (21) we have

$$\text{Prob}\{link\} = Q\left(\frac{10n\log_{10}(d/R_0)}{\sigma}\right) \quad (24)$$

where $Q(x) = \frac{1}{\sqrt{2\pi}}\int_x^\infty e^{-y^2/2}dy \quad (25)$

## V. NUMERICAL RESULTS AND DISCUSSIONS

In this section we present numerical results to get insight of sensing models and link model in WSN. The results are obtained by using MATLAB.

Fig. 5 shows the comparative study of network coverage versus number of nodes for different sensing models. We have shown coverage fraction for two different values of λ and σ. Curve-a is obtained for Boolean sensing model. Curves-c and f are obtained for λ = 0.01/m and 0.03/m respectively. It is clear from Fig. 5 that for higher value of λ, more number of node is required to provide a certain coverage fraction. Curves-b and d are obtained for σ = 2 dB and 8 dB respectively. It is also clear from the study that shadowing parameter degrades the network coverage. Curve-e is obtained for Elfes sensing model without approximation ($R_1$ = 10 m, λ = 0.03/m).

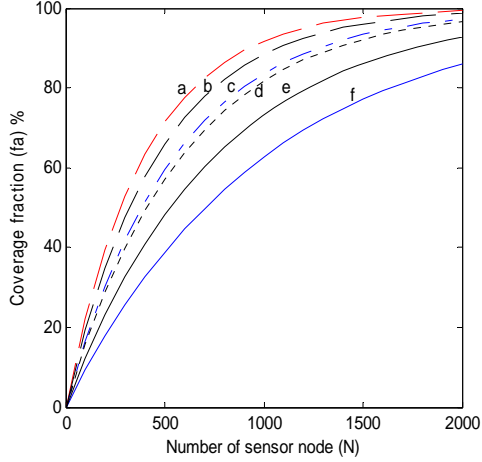

Fig. 5. Variation of coverage fraction for a circular area with radius 1000 m and maximum sensing radius $R_{max}$ = 50 m: (a) Boolean sensing, (b) Shadow fading (σ = 2 dB), (c) Elfes sensing (λ = 0.01/m), (d) Shadow fading (σ = 8 dB), (e) Elfes sensing without approximation ($R_1$ = 10 m, λ = 0.03/m), (f) Elfes sensing (λ = 0.03/m)

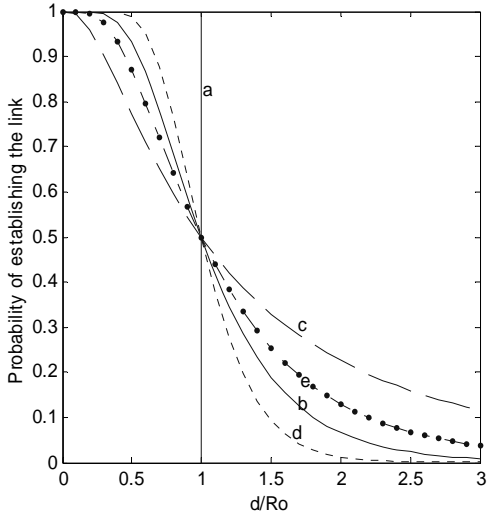

Fig. 6. Probability of establishing a link between two nodes: (a) σ = 0 dB; (b) n = 2, σ = 4 dB; (c) n = 2, σ = 8 dB; (d) n = 3, σ = 4 dB; (e) n = 3, σ = 8 dB.

Fig. 6 shows the variation of Probability of establishing a link between two nodes for different link distance. The link distance d is normalized to the communication range, $R_0$. It is clear that under the non-shadowed environment i.e. σ = 0 dB, there is no link beyond $R_0$. However, for other non-zero value of σ there is a finite probability of establishing a link when the link distance exceeds $R_0$. This comes at the cost of uncertainty in establishing a link when the link distance is less than $R_0$. This study implies that the shadowing effect improves the network connectivity.

## VI. CONCLUSIONS

In this paper network coverage and connectivity of WSN has been studied. It is clear from this study that individual sensing model has a great impact on network coverage. This study also reveals that shadow-fading parameters affect the coverage and connectivity. It is interesting to note that the shadowing effect reduces the network coverage while it enhances the connectivity of the network.